\documentclass{article}

\usepackage{arxiv}

\usepackage[utf8]{inputenc} 
\usepackage[T1]{fontenc}    
\usepackage[hidelinks]{hyperref}       
\usepackage{url}            
\usepackage{booktabs}       
\usepackage{amsfonts}       
\usepackage{nicefrac}       
\usepackage{microtype}      
\usepackage{lipsum}		
\usepackage{graphicx}
\usepackage{doi}

\usepackage{graphicx}
\usepackage{tablefootnote}

\renewcommand{\shorttitle}{\textit{arXiv} Template}

\title{Multi-style Training for South African \\ Call Centre Audio}

\date{}

\author{ 
    Walter~Heymans and Marelie H.~Davel \\
    Faculty of Engineering, \\
    North-West University, South Africa \\
    and CAIR, South Africa \\
	\texttt{walterheymans07@gmail.com} \\
	\And
	Charl van~Heerden \\
	Saigen, South Africa \\
	\texttt{charl@saigen.co.za} \\
}

\renewcommand{\shorttitle}{}

\hypersetup{
pdftitle={Multi-style Training for South African Call Centre Audio},
pdfsubject={cs.LG},
pdfauthor={Walter Heymans, Marelie H. Davel, Charl van Heerden},
pdfkeywords={Automatic speech recognition, Multi-style training, Call centre audio, WAV49 encoding},
}

\newcommand\blfootnote[1]{%
  \begingroup
  \renewcommand\thefootnote{}\footnote{#1}%
  \addtocounter{footnote}{-1}%
  \endgroup
}

\begin{document}
\maketitle
\blfootnote{This is a preprint - the final authenticated publication is available online at:\\ \url{https://doi.org/10.1007/978-3-030-95070-5_8}}

\begin{abstract}
Mismatched data is a challenging problem for automatic speech recognition (ASR) systems. One of the most common techniques used to address mismatched data is multi-style training (MTR), a form of data augmentation that attempts to transform the training data to be more representative of the testing data; and to learn robust representations applicable to different conditions. This task can be very challenging if the test conditions are unknown. We explore the impact of different MTR styles on system performance when testing conditions are different from training conditions in the context of deep neural network hidden Markov model (DNN-HMM) ASR systems. A controlled environment is created using the LibriSpeech corpus, where we isolate the effect of different MTR styles on final system performance. We evaluate our findings on a South African call centre dataset that contains noisy, WAV49-encoded audio.
\end{abstract}


\section{Introduction}
Automatic speech recognition (ASR) has been an active field of research since the 1970s and is still being developed and improved today~\cite{baker1975dragon, jelinek1975design, nassif2019speech}. The word error rate (WER), a typical measure of performance for ASR systems, has been significantly reduced over the last few decades. Main factors that contributed to this improvement were recent developments in deep learning, increased computational power of modern computers, specifically graphical processing units, and large amounts of collected data~\cite{lu2020automatic}.

ASR systems tend to perform more poorly when there is a large mismatch between training and testing data. Factors that contribute to this mismatch include various forms of background noise, microphone distortion, different recording environments, encoding noise, people that speak in different speaking styles and accents, etc. It is difficult for an ASR system to generalise to new audio with different conditions if no attempt is made during training to handle such variability in the data. 

A popular technique to address mismatch in audio for ASR is multi-style training (MTR)~\cite{ko2015audio, lippmann1987multi, li2012improving, doulaty2016automatic, park2019specaugment, szoke2019building}. MTR aims to transform the training data to be more representative of the testing data and to learn robust representations of the training data. A new training dataset is created from the existing set by adding a series of MTR styles using data augmentation. These can include: changing the speed and volume~\cite{ko2015audio}, speech style~\cite{lippmann1987multi} or sampling rate~\cite{li2012improving}; adding time and frequency distortions~\cite{park2019specaugment} or background noise; and simulating reverberation~\cite{szoke2019building}. The styles are typically chosen without knowledge of the testing conditions, and must still be able to handle a wide variety of mismatch. In addition, the number of styles that are added must be taken into consideration, because the computational cost of training an ASR system increases significantly with each style that is added.

In this work, we analyse the effects of MTR in a controlled environment using the LibriSpeech corpus~\cite{panayotov2015librispeech}. Speed, volume and noise perturbation are added to clean training data and evaluated on WAV49-encoded development and test sets. We show the performance gain as a result of individual and combined MTR styles. This provides a practical approach to improve ASR systems efficiently on WAV49-encoded audio, often used in South African call centres.

Section \ref{sec:related_work} gives a brief overview of related work in the field of MTR for deep neural network (DNN) based speech recognition. In Section \ref{sec:data}, we introduce the call centre dataset and how we created a controlled environment using the LibriSpeech corpus. Our experimental setup is explained in Section \ref{sec:network_arch} and the results are presented in Section \ref{sec:analysis}. Finally, the key findings are discussed in Section \ref{sec:conclusion}.

\section{Related Work}
\label{sec:related_work}
We are not aware of any studies that investigate MTR for DNN-based call centre ASR. There are studies that investigate the effects of different perturbation levels, but none of them focus on call centre audio. The perturbation types used in these studies include additive noise and room impulse responses~\cite{doulaty2016automatic}, speed \cite{ko2015audio} and volume \cite{gokay2019improving}.

Doulaty et al. investigated a method to automatically identify noise perturbation levels in a target set of utterances~\cite{doulaty2016automatic}. They used a `voice-search' dataset for their experiments. A noisy test set was created by perturbing a clean set with different perturbation styles including additive background noise and room impulse responses using different signal-to-noise ratios (SNRs). MTR was used to train a number of multi-layer perceptron (MLP) models, each with different perturbation levels. The MTR model that was trained on data with the closest matched conditions was then used to evaluate the target utterance. Their study revealed that accurately matched noise perturbation levels result in better ASR performance. This showed the importance of selecting conditions for MTR that are matched to that of the test set.

Speed perturbation is a common technique that is applied widely in MTR setups. Ko et al.~\cite{ko2015audio} investigated making two copies of the original dataset, one slowed down by 10\% and another made 10\% faster. An average relative improvement of 4.3\% was observed across 4 different tasks, with a relative improvement of 6.7\% on Switchboard \cite{godfrey1992switchboard}, a conversational telephone corpus. The improvement on the full LibriSpeech corpus \cite{panayotov2015librispeech} was 3.2\% and only 0.32\% relatively on the ASpIRE corpus \cite{harper2015automatic}. The authors attribute this small improvement on the ASpIRE corpus to simulated reverberation that was already applied to the training data.

Gokay and Yalcin investigated the effects of speed and volume perturbation in a low-resource 10 hour Turkish dataset of natural speech from a professional speaker~\cite{gokay2019improving}. They used an end-to-end ASR system based on Deep Speech 2~\cite{amodei2016deep}. Adding speed and volume perturbation, individually and together improved their WER by between 8.2\% and 12.9\% relatively. They also added 10 hours of new training data without MTR and the improvement was 26.3\% relatively, much better than any MTR technique they used. The effect of MTR was amplified because their original training dataset was very small. As more matched data is added, the improvements of MTR (speed and volume perturbation in this case) should become less apparent, because the training data itself can include more conditions.

Our work focuses on having a single model instead of multiple MTR models each trained with different conditions. We apply different perturbation styles (additive noise, speed and volume) to clean training data in a controlled environment to analyse the effect of each method on DNN-based ASR using the LibriSpeech corpus (encoded using WAV49 encoding). The findings are applied to a proprietary South African call centre dataset that is WAV49 encoded.

\section{Data}
\label{sec:data}
We use two datasets: our final aim is to use MTR to determine how much we can improve the performance of an ASR system on mismatched call centre data, but first experiment with individual styles in a controlled environment. An overview of the call centre dataset is given in Section \ref{sec:sa_call_centre}; and in Section \ref{sec:librispeech_corpus_wav49} we explain how the controlled environment is set up using the LibriSpeech corpus.

\subsection{South African Call Centre Dataset}
\label{sec:sa_call_centre}
Call centres handle very large amounts of data on a daily basis. Typically, all calls are recorded and stored for future reference, legal purposes and call centre speech analytics. Due to the large number of calls, the recordings are often compressed for longer term storage. This can decrease the required storage space by up to twenty times. Although compression is beneficial for storage requirements, it is a challenging problem for ASR systems. We use a proprietary South African call centre dataset, referred to from here as the SACC corpus. All data in the SACC corpus consists of narrow-band single channel recordings. The corpus is mostly South African English, but there are occasional non-English words from other official South African languages. Utterances with mostly non-English speech have been removed from the corpus. Table \ref{tab:call_centre_datasets} shows the datasets in the SACC corpus after mostly non-English utterances were removed. There are 48.8 hours of training data that were originally not encoded, with an encoded version created using Sox\footnote{\url{http://sox.sourceforge.net}}. The training, development and test sets were created from the same set by dividing the corpus into three parts. A 1.2 hour held-out test set that was recorded and processed by the call centre at a later stage is also available for final testing. Calls are compressed in three steps, namely, lowering the sampling rate, combining dual channel audio to a single channel, and encoding the audio with WAV49 encoding.

\subsubsection{Sampling Rate:}
Most high quality ASR systems work with wide-band audio that is sampled at 16 kHz. All data in the SACC corpus is narrow-band (8 kHz). Narrow-band audio has less frequency information available than wide-band, which tends to hurt ASR systems slightly.
 
\subsubsection{Channel Combination:}
Telephone calls usually have two channels, one for the call center agent and one for the client (person who called or is being called). Two audio channels use twice as much storage space as a single channel, which is why they are combined to form only one channel. This creates three problems for an ASR system: (1) noise from both channels are present in the new signal, (2) overlapping speech and (3) speaker confusion. All of these factors contribute to decreased speech recognition performance.

\subsubsection{WAV49 Encoding:}
Compressing audio with a codec can reduce the storage space significantly, but also keep most of the original audio quality. There are many different compression methods, such as: Free Lossless Audio Codec (FLAC), MPEG Audio Layer III (MP3), Advanced Audio Coding (AAC), Ogg Vorbis, Speex and Opus~\cite{siegert2016measuring}. The SACC corpus is WAV49-encoded: a full-rate GSM 06.10 codec with a compression ratio of 10:1 is applied to the audio file~\cite{etsi6300}. It is then saved in a WAV file format resulting in a WAV49-encoded file~\cite{van2019asterisk}.

\begin{table}[th]
\centering
\caption{SACC corpus subsets with sampling rate, encoding and total duration.}
\begin{tabular}[t]{l c c c}
    \hline
    \textbf{Dataset} & \textbf{Sampling rate} &  \hspace{0.25cm}\textbf{Encoding}\hspace{0.25cm} & \textbf{Hours} \\
    \hline
    train & 8 kHz & - & 48.8 \\
    train-e & 8 kHz & WAV49 & 48.8 \\
    \hline
    dev & 8 kHz & - & 7.1 \\
    dev-e & 8 kHz & WAV49 & 7.1 \\
    \hline
    test & 8 kHz & - & 6.2 \\
    test-e & 8 kHz & WAV49 & 6.2 \\
    \hline
    held-out test & 8 kHz & WAV49 & 1.2 \\
    \hline
\end{tabular}
\label{tab:call_centre_datasets}
\end{table}

\subsection{LibriSpeech Corpus with WAV49 Encoding}
\label{sec:librispeech_corpus_wav49}
The LibriSpeech corpus contains 1 000 hours of English audiobook recordings sampled at 16 kHz~\cite{panayotov2015librispeech}. It is a freely available public dataset that is used as a benchmark for many state-of-the-art ASR models~\cite{park2019specaugment,xu2021self}. The dataset contains about 460 hours of clean training data, 500 hours of noisy training data, two development and two test sets (one clean and one noisy each). There are also four different language models included ranging from a small tri-gram to a large unpruned 4-gram language model.

We use the LibriSpeech corpus to create a controlled environment for MTR experiments. The corpus is well suited for this, since a large portion of the corpus has been labeled as ``clean'', meaning that the recordings do not have much noise. Using these clean audio recordings, noise and encoding can easily be added to simulate call centre audio conditions.

We use the 100 hour subset of the LibriSpeech corpus for training data and the small tri-gram (tg-small) language model for faster decoding and making the comparison of different acoustic models more efficient. To simulate call centre conditions, we add background noise to the clean development (\textit{dev-clean}) and test (\textit{test-clean}) sets using the QUT-NOISE corpus~\cite{dean2010qut} with a signal-to-noise ratio (SNR) of 5 dB. For our training data, we add noise using the Musan noise corpus \cite{musan2015} to create an artificial mismatch in noise conditions. An artificial mismatch is created between training and test data, because the noise corpus used to create the test set has different types of noise than the corpus used to perturb the training data. For both noise corpora, we randomly add a noise file to each utterance for the total duration of the utterance. We also encode these sets with WAV49 encoding and reduce their sampling rate to 8 kHz using Sox.

Table \ref{tab:training_datasets_ms} shows the training datasets we use for MTR. Different combinations are used to see the combined performance impact on the development set. Speed and volume perturbation was applied to the training data with a change of 10\% and 20\% respectively. The speed was either increased or decreased with equal probability - approximately half of the utterances have a slower speed compared to the original set and the other half have a faster speed. Volume is handled in a similar way.
Table \ref{tab:test_datasets_ms} shows the development and test sets that are created from the \textit{dev-clean} and \textit{test-clean} subsets, in a similar manner as the training set but with fewer conditions.

\begin{table}[ht]
\centering
\caption{Multi-style training datasets created using the 100 hour clean LibriSpeech subset (\textit{train-clean-100}).}
\begin{tabular}[t]{l c c c c c}
    \hline
    \textbf{Dataset name} & \hspace{0.25cm}\textbf{Encoding}\hspace{0.25cm} & \textbf{Noise corpus} & \hspace{0.25cm}\textbf{SNR}\hspace{0.25cm} & \textbf{Speed} & \hspace{0.25cm}\textbf{Volume}\hspace{0.25cm} \\
    \hline
    train-clean & - & - & - & - & - \\
    train-clean-8k & - & - & - & - & - \\
    train-clean-e & WAV49 & - & - & - & - \\
    train-noisy-e-5 & WAV49 & QUT & 5 & - & - \\
    train-clean-e-s & WAV49 & - & - & 10\% & - \\
    train-clean-e-v & WAV49 & - & - & - & 20\% \\
    train-clean-e-sv & WAV49 & - & - & 10\% & 20\% \\
    train-musan-e-5 & WAV49 & Musan & 5 & - & - \\
    train-musan-e-10 & WAV49 & Musan & 10 & - & - \\
    train-musan-e-15 & WAV49 & Musan & 15 & - & - \\
    train-musan-e-20 & WAV49 & Musan & 20 & - & - \\
    train-musan-e-15-s & WAV49 & Musan & 15 & 10\% & - \\
    train-musan-e-15-v & WAV49 & Musan & 15 & - & 20\% \\
    train-musan-e-15-sv & WAV49 & Musan & 15 & 10\% & 20\% \\
    \hline
\end{tabular}
\label{tab:training_datasets_ms}
\end{table}

\begin{table}[ht]
\centering
\caption{Development and test datasets created using the LibriSpeech \textit{dev-clean} and \textit{test-clean} sets.}
\begin{tabular}[t]{l c c c c c}
    \hline
    \textbf{Dataset name}\hspace{0.25cm} & \textbf{Source dataset} & \hspace{0.25cm}\textbf{Encoding}\hspace{0.25cm} & \textbf{Noise corpus} & \hspace{0.25cm}\textbf{SNR}\hspace{0.25cm} & \textbf{Hours} \\
    \hline
    dev-clean-e & dev-clean & WAV49 & - & - & 5.4 \\
    dev-noisy-e-5 & dev-clean & WAV49 & QUT & 5 & 5.4 \\
    test-noisy-e-5 & test-clean & WAV49 & QUT & 5 & 5.4 \\
    \hline
\end{tabular}
\label{tab:test_datasets_ms}
\end{table}

\section{Experimental Setup}
\label{sec:network_arch}
We use the Pytorch-Kaldi\footnote{Available at: \url{https://github.com/mravanelli/pytorch-kaldi}} ASR toolkit to train a context-dependant deep neural network hidden Markov model (CD-DNN-HMM) ASR system~\cite{ravanelli2019pytorch}. We use the toolkit's default training setup with the standard scoring scripts for the LibriSpeech corpus. Everything in our setup is the same as theirs, except that we additionally optimise four selected hyperparameters (batch size, learning rate, language model weight and word insertion penalty).

We use the default MLP acoustic model for the LibriSpeech corpus. The acoustic model is a 5 hidden-layer network with 1 024 hidden units per layer. All hidden layers use rectified linear unit (ReLU) activation functions with batch normalisation and dropout with probability of 0.15. The output layer does not use batch normalisation or dropout and has a softmax activation function. We use feature-space maximum likelihood linear regression (fMLLR) input features with a temporal context window of 11 frames~\cite{gales1998maximum}. Our model trained on the \textit{train-clean-100} subset of the LibriSpeech corpus achieved similar results to public baselines~\cite{ravanelli2019pytorch}.

All networks are trained with the stochastic gradient descent (SGD) optimiser and a learning rate scheduler that halves the learning rate when the relative improvement\footnote{Senone error rate is used to measure performance after each training epoch.} on the development set is less than 0.001. The acoustic model is trained with the negative log-likelihood loss function to predict HMM state probabilities. The batch size and learning rate are optimised on the development set using a grid search. We found that these two hyperparameters have the largest effect on WER. The language model weight and word insertion penalty that gave the lowest WER on the development set are used for the final systems. All other hyperparameters are kept fixed. All networks are trained for 24 epochs; at this point all networks have converged. Each network is trained with three different random initialisation seeds. We report on the average WER and WER standard error across seeds.

\section{Analysis}
\label{sec:analysis}
In this section, we first investigate the effect of sampling rate differences between training and testing data on the LibriSpeech corpus (Section \ref{sec:sampling_rate_ls}). This is done to see how much of the performance is lost due to WAV49 encoding and how much is due to the difference in sampling rate. Then we create a controlled environment using the LibriSpeech corpus to isolate the effects that different styles in an MTR setup has on system performance (Section \ref{sec:mtr_ls}). Noise, speed and volume perturbation are applied to a clean training set to measure how much each of these techniques can improve the WER on the noise-perturbed test set. We also look at how larger networks can benefit MTR setups (Section \ref{sec:mtr_large_nets}). Finally, the findings in the controlled experiment are applied to the SACC corpus (Section \ref{sec:mtr_inhouse}).

\subsection{Sampling Rate Differences on the LibriSpeech Corpus}
\label{sec:sampling_rate_ls}
Two sets of networks are trained on the \textit{train-clean} and \textit{train-clean-e} datasets using 16 kHz and 8 kHz audio, respectively. Encoded data, which was previously downsampled during the encoding stage, was now upsampled to 16 kHz using Sox to train a 16 kHz model. We also downsampled the clean training and development sets to 8 kHz to measure the performance difference caused by WAV49 encoding when unencoded narrow-band audio is used for training. 

The WER results are shown in Table \ref{tab:results_encoding_sampling}. All data in the top section is used at 16 kHz (training and development sets); while the data in the bottom section is used at 8 kHz. The \textit{train-clean} model performed the best on the \textit{dev-clean} set for both frequencies. Evaluating the clean 16 kHz model (\textit{train-clean}, 16 kHz) on the encoded development set drastically increased the WER to 19.32\%. By downsampling the clean training data (\textit{train-clean}, 8 kHz), the result improved by 40.1\% relative WER to 11.44\%. This network, trained only using unencoded data, performed only slightly worse than the \textit{train-clean-e} model (11.44\% vs 11.21\% WER). The large difference in WER can be reduced significantly by downsampling unencoded training data. Encoding the training data only gave a slight improvement over the 8 kHz \textit{train-clean} model of 2.0\% relative WER.

We observe that most of the mismatch is a result of the difference in sampling rate and not due to encoding. The improvement achieved when downsampling unencoded training data was 40.1\% relative WER and only 2.0\% when encoding the training set.

\begin{table}[t]
\centering
\caption{WER results of models with different sampling rates on \textit{dev-clean} and \textit{dev-clean-e}. Average WER and standard error is shown over 3 seeds.}
\begin{tabular}[t]{l l c c}
    \hline
    \textbf{Train set} & \textbf{Sampling rate} & \hspace{0.25cm}\textbf{dev-clean}\hspace{0.25cm} & \textbf{dev-clean-e} \\
    \hline
    \textbf{Wide-band} & & & \\
    \hspace{0.25cm}train-clean & 16 kHz & \textbf{8.88 $\pm$ 0.10} & 19.32 $\pm$ 0.11 \\
    \hspace{0.25cm}train-clean-e\hspace{0.25cm} & 16 kHz\tablefootnote{Up-sampled from 8 kHz.} & 10.71 $\pm$ 0.05 & \textbf{11.02 $\pm$ 0.03} \\
    \hline
    \textbf{Narrow-band} & & & \\
    \hspace{0.25cm}train-clean & 8 kHz & 10.29 $\pm$ 0.03 & 11.44 $\pm$ 0.04 \\
    \hspace{0.25cm}train-clean-e\hspace{0.25cm} & 8 kHz & 10.76 $\pm$ 0.02 & 11.21 $\pm$ 0.03 \\
    \hline
\end{tabular}
\label{tab:results_encoding_sampling}
\end{table}

\subsection{Multi-style Training on the LibriSpeech Corpus}
\label{sec:mtr_ls}
In this experiment, we analyse the effect of different MTR styles on a set with mismatched noise conditions. Different combinations of noise, speed and volume perturbation are used for training data. 
Table \ref{tab:results_multistyle} shows the WER results on the development set (\textit{dev-noisy-e-5}). Information about the training datasets used by each model is shown in Table \ref{tab:training_datasets_ms}. All data, except for the 8 kHz \textit{train-clean-8k} set is upsampled to 16 kHz using Sox. The upsampling process used does not attempt to interpolate values in order to add high frequency information that was lost during downsampling.  Only the existing lower frequency components are retained.

\begin{table}[t]
\centering
\caption{WER on development set (\textit{dev-noisy-e-5}) using training datasets with different styles. Average WER and standard error is shown over 3 seeds.}
\begin{tabular}{l l c c}
    \hline
    \textbf{Model} & \textbf{Datasets} & \textbf{Size} & \textbf{Dev WER} \\
    \hline
    \textbf{Variations of clean set} \\
    \hspace{0.25cm}train-clean & train-clean & 1 & 36.46 $\pm$ 0.14 \\
    \hspace{0.25cm}train-clean-8k\tablefootnote{Training and test data is used at 8 kHz.} & train-clean @ 8 kHz & 1 & 33.23 $\pm$ 0.21 \\
    \hspace{0.25cm}train-clean-e & train-clean-e & 1 & \textbf{28.06 $\pm$ 0.03} \\
    \hline
    \textbf{Speed and volume} \\
    \hspace{0.25cm}train-clean-e-s & train-clean-e + s & 2 & \textbf{27.83 $\pm$ 0.02} \\
    \hspace{0.25cm}train-clean-e-v & train-clean-e + v & 2 & 28.21 $\pm$ 0.07 \\
    \hspace{0.25cm}train-clean-e-sv & train-clean-e + sv & 2 & 28.25 $\pm$ 0.10 \\
    \hspace{0.25cm}train-clean-e-s-v & train-clean-e + s + v & 3 & 28.04 $\pm$ 0.13 \\
    \hline
    \textbf{Noise} \\
    \hspace{0.25cm}train-musan-e-5 & train-musan-e-5 & 1 & 29.29 $\pm$ 0.34 \\
    \hspace{0.25cm}train-musan-e-10 & train-musan-e-10 & 1 & 27.29 $\pm$ 0.12 \\
    \hspace{0.25cm}train-musan-e-15 & train-musan-e-15 & 1 & \textbf{23.30 $\pm$ 0.06} \\
    \hspace{0.25cm}train-musan-e-20 & train-musan-e-20 & 1 & 26.64 $\pm$ 0.09 \\
    \hline
    \textbf{Speed, volume and noise} \\
    \hspace{0.25cm}train-musan-e-15-s & train-musan-e-15 + s & 2 & 24.09 $\pm$ 0.05 \\
    \hspace{0.25cm}train-musan-e-15-v & train-musan-e-15 + v & 2 & 23.97 $\pm$ 0.11 \\
    \hspace{0.25cm}train-musan-e-15-sv & train-musan-e-15 + sv & 2 & 24.34 $\pm$ 0.02 \\
    \hspace{0.25cm}train-musan-e-15-s-v & train-musan-e-15 + s + v & 3 & \textbf{23.80 $\pm$ 0.09} \\
    \hspace{0.25cm}train-musan-e-15-s-v-sv & train-musan-e-15 + s + v + sv & 4 & 23.89 $\pm$ 0.08 \\
    \hline
    \textbf{Matched noise} \\
    \hspace{0.25cm}train-noisy-e-5 & train-noisy-e-5 & 1 & \textbf{19.75 $\pm$ 0.04} \\
    \hline
\end{tabular}
\label{tab:results_multistyle}
\end{table}

Similar to the results in Section \ref{sec:sampling_rate_ls}, downsampling the clean training data improved the WER, although the improvement is much less than before. Adding WAV49 encoding to the training data improved the relative WER of the \textit{train-clean-8k} model by 15.6\%, much more than observed in Section \ref{sec:sampling_rate_ls}.

Speed and volume perturbation were applied to the clean encoded training data. Different combinations of datasets were evaluated. A small improvement in WER (0.8\% relative) was observed when using only speed perturbation. None of the other combinations resulted in a notable improvement; the \textit{train-clean-e-v} and \textit{train-clean-e-sv} networks performed worse than without perturbation. This may be because the training data already captures a large range of speed and volumes, or that the development set does not vary much in terms of speed and volume. 

Noise perturbation was applied to clean training data using four different SNR values and the sets were encoded afterwards. The performance with the 15 dB network was much better than the rest (12.5\% to 20.4\% relative WER), despite the fact that the development set used an SNR of 5 dB. The different noise corpora, QUT-NOISE vs Musan, can explain the difference. Energy in the noise files are distributed differently, so the SNR values are not directly comparable. The \textit{train-musan-e-15} model performed 17.0\% relatively better than the \textit{train-clean-e} model. This emphasises how important matched training and test conditions are. It is very important to use the correct SNR for noise perturbation, because it has a large influence on system performance.

We used the best noise-perturbed training dataset and added speed and volume perturbation. The WER when using speed and volume perturbation in any combination did not improve the result compared to using only additive noise. A similar phenomenon was observed in \cite{ko2015audio} on the ASpIRE corpus, where they only observed an absolute WER improvement of 0.1\%. We further investigate this in Section \ref{sec:mtr_large_nets}.

Finally, we trained a network using a noise-matched training dataset, \textit{train-noisy-e-5} that also uses the QUT-NOISE corpus. The WER for this model is 17.0\% relatively better than the best MTR model. This shows that MTR has many shortcomings when test conditions are significantly different from the training set. When encountering unseen environments on a new test set, most systems will probably struggle to do well.

Up to this point, all results were reported on the development set. Table \ref{tab:results_multistyle_test} shows the WERs on the test set (\textit{test-noisy-e-5}) using six selected models that performed the best in each category on the development set. The results on the test set are very similar to those on the development set. The best MTR model of all the combinations tested, is the one that used only noise perturbation using the SNR value that performed the best on the development set. The difference in WER between the best MTR model and the \textit{train-clean} model is 33.5\% relative. MTR can clearly reduce the WER significantly if the conditions are properly chosen, but MTR still performed 16.5\% worse than the \textit{train-noisy-e-5} model with matched conditions.

\begin{table}[tb]
\centering
\caption{WER on test set (\textit{test-noisy-e-5}) using training datasets with different styles. Average WER and standard error is shown over 3 seeds.}
\begin{tabular}{l l c c}
    \hline
    \textbf{Model} & \textbf{Datasets} & \textbf{Size}\hspace{0.25cm} & \textbf{Test WER} \\
    \hline
    \textbf{Variations of clean set} \\
    \hspace{0.25cm}train-clean & train-clean & 1 & 36.92 $\pm$ 0.18 \\
    \hspace{0.25cm}train-clean-e & train-clean-e & 1 & 29.16 $\pm$ 0.12 \\
    \hline
    \textbf{Speed and volume} \\
    \hspace{0.25cm}train-clean-e-s & train-clean-e + s & 3 &  28.61 $\pm$ 0.11\\
    \hline
    \textbf{Noise} \\
    \hspace{0.25cm}train-musan-e-15 & train-musan-e-15 & 1 & \textbf{24.54 $\pm$ 0.22} \\
    \hline
    \textbf{Speed, volume and noise} \\
    \hspace{0.25cm}train-musan-e-15-s-v & train-musan-e-15 + s + v & 3 & 24.87 $\pm$ 0.08 \\
    \hline
    \textbf{Matched noise} \\
    \hspace{0.25cm}train-noisy-e-5 & train-noisy-e-5 & 1 & \textbf{20.48 $\pm$ 0.03} \\
    \hline
\end{tabular}
\label{tab:results_multistyle_test}
\end{table}

\subsection{Multi-style Training Using Larger Networks on the LibriSpeech Corpus}
\label{sec:mtr_large_nets}
By adding speed and volume perturbation, you also add more training data. It is possible that the network with only 1 024 hidden units is too small to capture the larger data distribution. Increasing the network capacity should help the models using more training datasets generalise better and possibly give an advantage to the speed and volume perturbation networks.

Using the same training and optimisation protocol described in Section \ref{sec:network_arch}, we train three networks with 2 048 hidden units per layer instead of 1 024. The average WERs are shown in Table \ref{tab:results_large_nets} over three seeds for both network sizes; the dimensions of the hidden layers are shown in brackets. The performance of all models improved, but the model using speed and volume perturbation improved more than the model only using noise perturbation. The larger capacity benefits the model with more training data, but also the model using only encoded training data. The difference between the small and large network for the \textit{train-musan-e-15} model is almost negligible.

\begin{table}[t]
\centering
\caption{WER on development  (\textit{dev-noisy-e-5}) and test set (\textit{test-noise-e-5}) using MLP acoustic models with 2 048 hidden units per layer. Average WER and standard error is shown over 3 seeds.}
\begin{tabular}{l c c c}
    \hline
    \textbf{Model} & \hspace{0.25cm}\textbf{Size}\hspace{0.25cm} & \hspace{0.25cm}\textbf{Dev WER}\hspace{0.25cm} & \hspace{0.25cm}\textbf{Test WER}\hspace{0.25cm} \\
    \hline
    \textbf{Encoded} \\
    \hspace{0.25cm}train-clean-e (1 024x5) & 1 & 28.06 $\pm$ 0.03 & 29.16 $\pm$ 0.12 \\
    \hspace{0.25cm}train-clean-e (2 048x5) & 1 & 26.82 $\pm$ 0.08 & 27.74 $\pm$ 0.14 \\
    \hline
    \textbf{Noise} \\
    \hspace{0.25cm}train-musan-e-15 (1 024x5) & 1 & 23.30 $\pm$ 0.06 & 24.54 $\pm$ 0.22 \\
    \hspace{0.25cm}train-musan-e-15 (2 048x5) & 1 & 23.15 $\pm$ 0.10 & 24.55 $\pm$ 0.04 \\
    \hline
    \textbf{Speed, volume and noise} \\
    \hspace{0.25cm}train-musan-e-15-s-v (1 024x5) & 3 & 23.80 $\pm$ 0.09 & 24.87 $\pm$ 0.08 \\
    \hspace{0.25cm}train-musan-e-15-s-v (2 048x5) & 3 & \textbf{22.81 $\pm$ 0.06} & \textbf{24.34 $\pm$ 0.08} \\
    \hline
\end{tabular}
\label{tab:results_large_nets}
\end{table}

This experiment confirmed the hypothesis that the MTR models required more capacity to outperform the noise-perturbed network. There is however an increased computational cost when doubling the number of hidden units on top of the three times more training data. This becomes an important trade-off to consider if computational resources are limited.

\subsection{Multi-style Training on the SACC Corpus}
\label{sec:mtr_inhouse}
We now evaluate our findings on the SACC corpus described in Section \ref{sec:sa_call_centre}. A baseline DNN acoustic model is trained using fMLLR features and 8 kHz unencoded training data using the protocol described in Section \ref{sec:network_arch}. Another network is trained using only a single set of WAV49-encoded training data. For MTR, we use normal and encoded training data, and apply speed perturbation to the encoded training data only. We did not include volume perturbation, because the experiment on the LibriSpeech corpus did not show a consistent improvement when using it. The unencoded training data is included, because we want to jointly perform well on both encoded and unencoded testing data. We did not add any noise, because the training data was already noisy and came from the same call centre as our development and test sets. The results on the development and test sets are shown in Table \ref{tab:call_centre_results}.

The model that used only the encoded training set performed better on encoded test sets, but similar to the baseline model on the \textit{dev} set and worse on the \textit{test} set. The MTR model performed the best across all five datasets with relative WER improvements of between 1.1\% and 2.9\%.

\begin{table}[bt!]
\centering
\caption{WER results on dev/test sets for the SACC corpus. Average WER is shown over 3 seeds. }
\begin{tabular}[t]{l c c c c c}
    \hline
    \textbf{Model} & \textbf{dev} & \hspace{0.25cm}\textbf{dev-e}\hspace{0.25cm} & \hspace{0.25cm}\textbf{test}\hspace{0.25cm} & \hspace{0.25cm}\textbf{test-e}\hspace{0.25cm} & \textbf{held-out test} \\
    \hline
    train & 28.41 & 28.91 & 33.14 & 33.43 & 41.90 \\
    train-e\hspace{0.25cm} & 28.40 & 28.63 & 33.36 & 33.04 & 41.80 \\
    MTR & \textbf{27.98} & \textbf{28.19} & \textbf{32.77} & \textbf{32.46} & \textbf{41.42} \\
    \hline
\end{tabular}
\label{tab:call_centre_results}
\end{table}

Since the training data is well matched with the testing data, MTR does not provide large improvements. It does not hurt performance, and small consistent improvements are possible, but the real advantage of MTR is only observed if there is a significant mismatch.

\section{Conclusion}
\label{sec:conclusion}
MTR can have a very positive effect on ASR performance, given that the styles are chosen appropriately. Speed and volume perturbation can slightly reduce WER in some scenarios, but are computationally much more expensive. When using a system on narrow-band test sets, training with narrow-band audio is absolutely necessary. 

The two styles that gave the best improvement on the WAV49 encoded LibriSpeech corpus was: (1) encoding training data (2) and noise perturbation if there is no noise in the training set. However, if the SNR of the added noise is completely different to the test conditions, it can hurt the system. The best MTR setup outperformed the clean baseline on the test set by 33.5\% relative WER, but still performed worse than the noise-matched model by 16.5\%.

Only two MTR styles were used on the SACC corpus (encoding and speed perturbation), since the recordings already contained noise and were narrow-band. The relative WER improvement on the test sets were limited, ranging from 1.1\% to 2.9\% when using MTR. The improvements of MTR are small on this corpus, which we attribute to the training and test data being well matched. 

With proper network capacity, MTR does not hurt system performance, even when the data is very well matched. Consistent small improvements are observed in matched datasets, with very large improvements achieved on mismatched datasets.

\section*{Acknowledgement}
The authors acknowledge the Centre for High Performance Computing (CHPC), South Africa, for providing computational resources used in this research.

\bibliographystyle{unsrt}
\bibliography{references}

\end{document}